\documentclass{ws-procs975x65}

\begin{document}
\def\lsim{\lower.5ex\hbox{$\; \buildrel < \over \sim \;$}}
\def\gsim{\lower.5ex\hbox{$\; \buildrel > \over \sim \;$}}
\def \simeq{\lower.3ex\hbox{$\; \buildrel \sim \over - \;$}}
\def\ch{\lower-0.55ex\hbox{--}\kern-0.55em{\lower0.15ex\hbox{$h$}}}
\def\lh{\lower-0.55ex\hbox{--}\kern-0.55em{\lower0.15ex\hbox{$\lambda$}}}
\def\eg{{\it e.g.,} }
\def\etal{{\em et al.} }
\def\ie{{\em i.e.,} }
\def\ee{{$e^--e^+$}}
\def\ep{{$e^--p^+$}}

\title{Accretion of multi-species plasma onto black holes}

\author{Indranil Chattopadhyay $^*$} 

\address{ARIES, Manora Peak,\\
Nainital-263129, Uttarakhand, India\\
$^*$E-mail: indra@aries.res.in\\
www.aries.res.in}

%
%
\begin{abstract}
Matter falling onto a black hole is trans-relativistic, transonic, and close to the horizon it is sub-Keplerian. Such a flow shows the existence of multiple critical point, shocks etc. Employing
relativistic equation of state and realistic composition, it has been shown that the solution strongly depend on the composition of the fluid. Electron-positron fluid is the least relativistic, to the extent that multiple critical points, shocks etc do not form. The most relativistic fluid is the one whose ratio of proton to electron number density is $\sim 0.2$.
Since the solution strongly depend on composition, the emitted radiation should strongly depend on
the composition too. 
\end{abstract}


\bodymatter

\section{Assumptions and governing equations}\label{aba:sec1}
According to general relativity, matter accreting onto a black hole will be trans-relativistic, transonic and sub-Keplerian before crossing the horizon.
Such kind of flow has observational verification
too \cite{shms01,shs02}. For transonic-rotating flow, 
the variation in the relative strength between gravity
and rotation,
causes the formation of multiple critical points (\ie $r_c$), in a significant part
of the energy-angular momentum parameter space \cite{lt80}. One of the consequences of
multiple $r_c$, is that matter flowing through one of the $r_c$ may jump onto the solution which passes through another $r_c$, \ie shock transition \cite{f87,c89,c96}.
Shock in accretion has been held responsible to supply the hot electrons to produce
hard power-law tail, power jets etc \cite{cm06,mc08,cd07,detal09}.
Therefore the study of the dynamics of the sub-Keplerian component of the accretion disc is very important.
It is generally believed that the flow solutions do not depend on the composition of the fluid. This is correct only for non-relativistic
($T<10^7$K) and ultra-relativistic ($T\gsim 10^{13}K$) temperatures, but for $10^7$K$\lsim T\lsim 10^{13}$K, flow solutions
completely depend on the composition \cite{cr09}, to the extent that the electron-positron fluid ($\xi=0$ or \ee)
is the least relativistic. The question is, how would fluid composition affect the solutions of rotating flow?

The equations of motion of relativistic fluid are $T^{\mu \nu}_{;~\nu}=0$ and $(nu^{\nu})_{;~\nu}=0$,
and the fluid is assumed to be in steady state, axis-symmetric and is in vertical equilibrium. The energy density or equation of state is given by,\cite{cr09}
\begin{equation}
e=n_{e^-}m_{e^-}c^2f,
\end{equation}
where, 
\begin{equation}
f=(2-\xi)\left[1+\Theta\left(\frac{9\Theta+3}{3\Theta+2}\right)\right]
+\xi\left[\frac{1}{\eta}+\Theta\left(\frac{9\Theta+3/\eta}{3\Theta+2/\eta}\right)\right].
\end{equation}
here, $\xi=n_{p^+}/n_{e^-}=$ratio between proton and electron number densities, $\eta=m_{e^-}/m_{p^+}=$
ratio between electron and proton mass, and $\Theta=kT/m_{e^-}c^2$. The equations of motion are solved by the standard technique
of critical point analysis \cite{c90}. 
\begin{figure}[t]
\begin{center}
\psfig{file=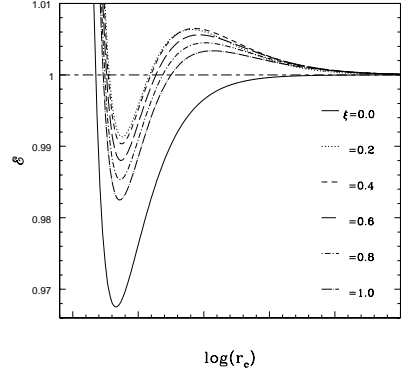,width=2.25in}
\end{center}
\caption{${\cal E}-{\rm log}(r_c)$ for $\xi=0.0$ (solid), $0.2$ (dotted), $0.4$ (dashed), $0.6$ (long dashed),
$0.8$ (dashed-dotted) and $1.0$ (long dashed-dotted). The long-short dashed curve is the bound energy level. For all curves $\lambda=3.2$.
}
\label{aba:fig1}
\end{figure}
In Fig. 1, the specific energy ${\cal E}$ ($=h~u_t$; $h$ is the specific enthalpy) is plotted as a function of the $r_c$. The composition or $\xi$ is marked on the figure, and all the curves are plotted for specific angular momentum $\lambda =3.2$. It is evident that, multiple $r_c$ do not exist for \ee fluid. Multiplicity of $r_c$ is achieved for fluids with $\xi \neq 0$. Moreover, at same $r_c$, ${\cal E}$ is maximum for $\xi \sim 0.2$, and minimum for \ee fluid.
Therefore, actual solutions should depend very strongly on ${\cal E}$, $\lambda$ and $\xi$. 
\begin{figure}
\begin{center}
\psfig{file=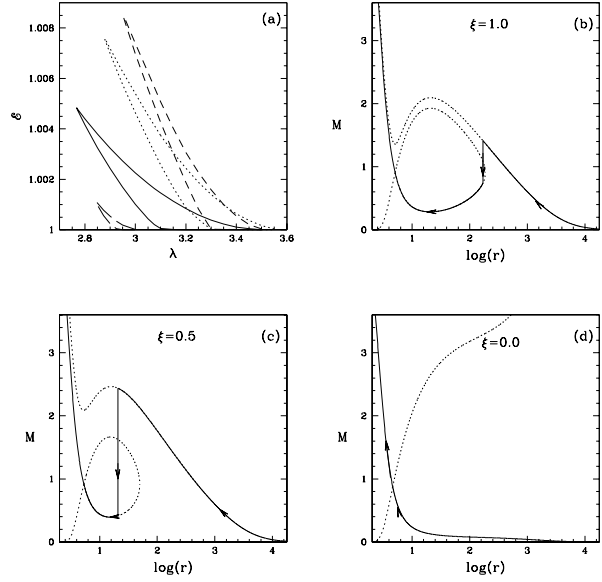,width=3.25in}
\end{center}
\caption{(a) The portion of ${\cal E}-\lambda$ parameter space which admits shock in accretion for $\xi=1.0$ (solid), $0.6$ (dotted), $0.2$ (dashed), $0.025$ (long dashed). Mach number $M$ variation
of accretion solution (solid and arrowed plot) for (b) $\xi=1.0$, (c) $\xi=0.5$ and (d) $\xi=0.0$
for the same set of parameter $\{{\cal E},\lambda\}=\{1.00028,3.3\}$
}
\label{aba:fig2}
\end{figure}

In Fig. 2a, the bounded region of ${\cal E}-\lambda$ parameter space, admits shock. The ${\cal E}-\lambda$ shock-domain for electron proton fluid (solid),
is remarkably different from that of a fluid with $\xi=0.6$ (dotted) or $\xi=0.2$ (dashed) or $\xi=0.025$ (long dashed). The shock domain shifts to the right as $\xi$ is decreased in the range
$1\geq \xi \gsim 0.2$ . Further decrease of $\xi$ in the range $0.2 \gsim \xi >0$, causes the bounded area to shrink and a shift towards left. The bounded area is zero for $\xi=0$. 
In Figs. 2b-2d, we present solutions of fluid of different $\xi$ but for same $\{{\cal E},\lambda\}=\{1.00028,3.3\}$. At this parameter, electron-proton fluid ($\xi=1$)
showed a very weak shock situated at a distance $169.19~GM/c^2$, but a fluid with $\xi=0.5$
with same parameters showed a strong shock at $20.68~GM/c^2$. While \ee fluid has only one critical point close to the horizon and showed no shock transition. Infact Fig. 2a showed that no shock can form in an \ee fluid.

\section{Conclusion}
The accretion solutions around a black hole not only depends on the energy, angular  momentum or viscosity parameters, but also on the composition. In particular, \ee fluid do not show the properties
of hot accreting flow, for \eg multiplicity of critical point or shock. Since the flow solutions depends strongly on composition, then the spectrum will depend strongly on composition too.

\end{document}